\documentclass[12pt]{article}
 
\def\be{\begin{equation}} 
\def\ee{\end{equation}} 
\def\cos{\mbox{cos}}
\def\sin{\mbox{sin}}
\def\exp{\mbox{exp}}
\def\sign{\mbox{sign}}
\def\det{\mbox{det}}
\def\a{\alpha}
\def\b{\beta}
\def\la{\langle}
\def\ra{\rangle}

\begin{document}

\begin{center}
{\large Fisher-Hartwig conjecture and the correlators in the 
inpenetrable Bose gas.}
\end{center}

\begin{center}
{\large A.A.Ovchinnikov}
\end{center}

\begin{center}
{\it Institute for Nuclear Research, RAS, Moscow, 117312, Russia}  
\end{center}

\vspace{0.1in}

\begin{abstract}

We apply the theorems from the theory of Toeplitz determinants 
to calculate the asymptotics of various correlators including 
the exponential ones in the inpenetrable one-dimensional Bose gas system. 
The known correlators in the free-fermion system are also used to 
test the generalized Fisher-Hartwig conjecture.

\end{abstract}

\vspace{0.3in}

{\bf 1. Introduction} 

\vspace{0.1in}

At present time the calculation of the correlators in the 
one-dimensional exactly solvable models including the 
constant for the asymptotic power-law behaviour remains an
open problem. Except the correlators in the XY - spin chain 
and the Ising model the calculations are possible for the 
case of one-dimensional Bose gas with $\delta$ - function 
interaction \cite{LL} in the strong coupling limit \cite{L}. 
In this case as well as in the case of the free-fermion 
system, the correlators are reduced to the Toeplitz 
determinants \cite{L} 
which allows to calculate their asymptotics exactly. 

In the present letter to calculate the asymptotics of the 
determinants for both systems 
we use the conjecture from the theory of Toeplitz 
matrices, \cite{BM}, the generalized Fisher-Hartwig conjecture,   
which is based on the proofs of the original 
Fisher-Hartwig conjecture \cite{FH} including the constant in 
front of the asymptotics, in a number of the particular cases 
\cite{Basor}, \cite{ES} (see also \cite{BM} and references 
therein). 
    One finds the representation of the generating function 
$f(x)$ of the Toeplitz matrix 
$M_{ij}=M(i-j)=\int_{0}^{2\pi}(dx/2\pi)e^{i(i-j)x}f(x)$ 
of the following form: 
\be
f(x)=f_0(x)\prod_{r}e^{ib_r(x-x_r-\pi\sign(x-x_r))}
(2-2\cos(x-x_r))^{a_r},  
\label{f}
\ee
where $x\in(0,2\pi)$ is implied, the discontinuities (jumps and zeroes 
or the power-law singularities) are at finite number of the points $x_r$,  
and $f_0(x)$ is the smooth non-vanishing function 
with the continuously defined argument at the interval $(0,2\pi)$.  
The function (\ref{f}) is 
characterized by the parameters $a_r$, $b_r$ at each point of  
discontinuity $x_r$. 
In general there are several representations of the form (\ref{f}) 
for a given functions $f(x)$. 
To obtain the asymptotics of the determinant one should take the sum 
over the representations (\ref{f}) corresponding to the minimal exponent 
$\sum_r(b_r^2-a_r^2)$: 
\be
D(N)=\sum_{Repr.}e^{l_0 N}N^{\sum_r(a_r^2-b_r^2)} E
\label{DN}
\ee
where $E$ is the constant independent of $N$, 
\be
E=\exp\left( \sum_{k=1}^{\infty}k l_k l_{-k}\right) 
\prod_{r}(f_{+}(x_r))^{-a_r+b_r}(f_{-}(x_r))^{-a_r-b_r}~~~~~~~~~~~~
\label{E}
\ee
\[
~~~~~~~~~\prod_{r\neq s}\left(1-e^{i(x_s-x_r)}\right)^{-(a_r+b_r)(a_s-b_s)}  
\prod_{r}\frac{G(1+a_r+b_r)G(1+a_r-b_r)}{G(1+2a_r)},  
\]
$l_k=\int_{-\pi}^{\pi}(dx/2\pi)e^{ikx}\ln(f_0(x))$, and 
the functions $f_{\pm}(x)$ are given by the equations  
\[
\ln f_{+}(x)=\sum_{k>0}l_{-k}e^{ikx}, ~~~
\ln f_{-}(x)=\sum_{k>0}l_{k}e^{-ikx},   
\]
where $G$ is the Barnes $G$-function \cite{Barnes}, 
$G(z+1)=G(z)\Gamma(z)$, $G(1)=1$. 
In ref.\cite{Basor} the theorem was proved for $a_r=0$ and 
an arbitrary number of the discontinuities of the imaginary part 
of the magnitude less than $1/2$, $|b_r|<1/2$, 
however, there are many reasons to believe it to be true also in 
the case $|b_r|=1/2$ \cite{BM}.  
For the case of an arbitrary single Fisher-Hartwig singularity 
the conjecture (\ref{DN}) was recently proved in ref.\cite{ES}.  
See ref.\cite{BM} for the complete list of the cases for which 
the rigorous proof of the conjecture (\ref{DN}) is available. 
      Schematically the rigorous proofs in the particular 
cases \cite{Basor} \cite{ES} go as follows.  
Suppose that equation (\ref{DN}) is fulfilled for some functions 
$f_1(x)$ and $f_2(x)$ of the class of piecewise continuous 
functions with continuously defined argument. 
Then the equation (\ref{DN}) is fulfilled for the function 
$f(x)=f_1(x)f_2(x)$. 
Thus it is sufficient to prove eq.(\ref{DN}) for the 
smooth function with the continuously defined argument, 
in which case it is reduces to the well known 
strong Szego theorem and for the singular function of the 
form (\ref{f}) $f(x)=(1-z)^{\a}(1-1/z)^{\b}$, 
$z=e^{ix}$, in which case the asymptotics is known exactly. 

The goal of the present letter is two-fold. 
First, the calculation of the correlators in the 
one-dimensional Bose gas and the free-fermion system 
provide the additional tests of the generalized Fisher-Hartwig 
conjecture \cite{BM} in the cases when the rigorous proofs 
are not available. Second, we calculate the asymptotics of 
the two kinds of the exponential correlators which is 
interesting by itself since they can be compared with the 
predictions of various approaches to the calculation of the 
asymptotics of the correlation functions in the exactly 
solvable models such as the harmonic fluid 
approach \cite{H} or the conformal field theory approach 
\cite{C}.

\vspace{0.2in}

{\bf 2. Calculation of the correlators.}

\vspace{0.1in}

  The calculation of the correlators in the inpenetrable 
Bose gas or the free fermion systems are based on the representation 
of the Toeplitz determinant as a random matrix average over the unitary 
group. It is easy to prove the following equation \cite{L}: 
\be
\det_M(M_{ij})=\frac{1}{M!}\prod_{i=1}^{M}\int_{0}^{2\pi}\frac{dx_i}{2\pi}
f(x_i)\prod_{i<j}|z_i-z_j|^{2}, ~~~~~~z_i=e^{ix_i},  
\label{rep}
\ee
where the function $f(x)$ corresponds to the matrix $M_{ij}=M(i-j)$. 
In fact, we have the following simple chain of equations: 
\[
\prod_{i=1}^{M}\int_{0}^{2\pi}\frac{dx_i}{2\pi}
f(x_i)\prod_{i<j}|z_i-z_j|^{2}=
\prod_{i=1}^{M}\int_{0}^{2\pi}\frac{dx_i}{2\pi}f(x_i)
\sum_{P,P^{\prime}}(-1)^{P+P^{\prime}}e^{i\sum_{i}x_i(Pi-P^{\prime}i)}= 
\]
\[
\sum_{P,P^{\prime}}(-1)^{P+P^{\prime}}\prod_{i}M(Pi-P^{\prime}i)= 
M!\det\left(M_{ij}\right), 
\]
where $P,P^{\prime}\in S_M$ are the permutations. 
Since the ground state wave functions for the free fermion system and 
the inpenetrable Bosons systems can be represented as 
\be
\psi_{ff}(x_1\ldots x_M)=\prod_{i<j}\sin(\pi x_{ij}/L), ~~~~ 
\psi_{b}(x_1\ldots x_M)=\prod_{i<j}|\sin(\pi x_{ij}/L)|,   
\label{wf}
\ee
where $x_{ij}=x_i-x_j$ and $L$ is the length of the system, 
the correlators are represented as Toeplitz 
determinants according to the formula (\ref{rep}).

We begin with the the following equal-time correlation function 
defined in terms of the Bose creation and annihilation operators 
$\phi^{+}(x)$, $\phi(x)$: 
\be
G_{\a}(x)= \la 0|\phi^{+}(x)e^{i\a N(x)}\phi(0)|0\ra, 
\label{ga}
\ee
where $\a$ is an arbitrary parameter and $N(x)$ is the operator 
of the number of particles at the segment $(0,x)$. In the framework 
of the first-quantization this correlator has the following form: 
\be
G_{\a}(x)= M\prod_{i=2}^{M}\int_{0}^{L}dx_i e^{i\a\sum_{i=2}^{M}\theta(x-x_i)}
\psi(x,x_2,...x_M)\psi(0,x_2,...x_M), 
\label{gga}
\ee
where the wave functions are normalized to unity and correspond to Bose-
statistic. One can see that at $\a=0$ the correlator (\ref{gga}) is the 
one-particle density matrix for the inpenetrable bosons system, 
while at $\a=\pi$ the correlator is the equal-time Green function 
(one-particle density matrix) of the free-fermion system.

In fact, in this case the integrand in eq.(\ref{gga}) reduces to the 
product of two free-fermion ground state wave functions 
(one should take into account the equation 
$\sin(\pi(y-x)/L)=\sign(y-x)(1/2)(2-2\cos(2\pi(y-x)/L))^{1/2}$, $0<y<L$). 
   First, we calculate the correlator (\ref{gga}) at $\a<\pi$.  
Clearly, taking into account the normalization of the wave functions (\ref{wf}),  
the equation $|\sin(\pi x/L)|=(1/2)(2-2\cos(2\pi x/L))^{1/2}$, and the equation 
\[
|\sin(\pi x_{ij}/L)|^2=(1/4)|e^{i2\pi x_i/L}-e^{i2\pi x_i/L}|^2, 
\]
one finally obtains the determinant of the form (\ref{rep}): 
\[
G_{\a}(x)=\frac{1}{L}\det_{M-1}M_{ij}[f(y)]
\]
where the function $f(y)$ ($0<y<2\pi$) equals 
\be
f(y)=(e^{i\a};1)(y)(2-2\cos(y-2\pi x/L))^{1/2}(2-2\cos(y))^{1/2}, 
\label{fy}
\ee
where we denote $(e^{i\a};1)(y)=e^{i\a}\theta(y-x_r)+\theta(x_r-y)$, 
$x_r=2\pi x/L$. 
This function should be represented in the form (\ref{f}) which gives 
\be
f_{0}(y)=e^{ibx_r}=e^{i\a x/L}, ~~~x_1=0, ~~a_1=\frac{1}{2}, ~~b_1=-b, ~~
x_2=x_r, ~~a_2=\frac{1}{2}, ~~b_2=b, 
\label{ab}
\ee
where $b=\a/2\pi<1/2$. 
Substituting this function into the determinant (\ref{DN}) we finally 
obtain the following expression for the correlator:
\[
G_{\a}(x)= e^{i\a(Mx/L-1/2)}G^{2}\left(\frac{3}{2}+b\right)
G^{2}\left(\frac{3}{2}-b\right)
\rho\frac{1}{\left(2M\sin\left(\pi x/L\right)\right)^{1/2+2b^2}}, 
\]
where $b=\a/2\pi<1/2$ and the particle density $\rho=M/L$. 
Note that the correlator (\ref{ga}) is the periodic function of 
the parameter $\a$ since at $\a>\pi$ $(b>1/2)$ one should use the 
function of the form (\ref{f}) with the parameters $b_r=\pm(1-b)$ 
in order to have the minimal exponent in the equation (\ref{DN}). 
From this equation at $b=0$ we obtain the asymptotics 
of the one-particle density matrix for the inpenetrable Bose gas \cite{L}: 
\[
G(x)=\rho\frac{1}{\left(2M\sin\left(\pi x/L\right)\right)^{1/2}}
G^{4}(3/2).   
\]
 
Now let us turn to the calculation of the free-fermion Green function 
which corresponds to the value $\a=\pi$ or $b=\a/2\pi=1/2$. 
In contrast to the case $\a<\pi$ in this case there are two different 
representations of the function (\ref{fy}) in the form (\ref{f}) which
corresponds to the equal value of the exponent $\sum_r(a_r^2-b_r^2)$ in 
the asymptotics (\ref{DN}). Namely, we have the same parameters as in 
eq.(\ref{ab}) except the two possible choises: $b_1=-b_2=1/2$, 
$b_1=-b_2=-1/2$. Taking the sum of this two terms we finally obtain the 
correct expression  
\[
G_{ff}(x)=\frac{\sin(\pi Mx/L)}{L\sin(\pi x/L)}, 
\]
which provides an important test of the generalized Fisher-Hartwig 
conjecture (\ref{E}).

Now we calculate the exponential correlator $\la0|e^{i\a N(x)}|0\ra$, 
where again the operator $N(x)$ is the operator of the number of 
particles at the segment $(0,x)$. In the first- quantization 
language this correlator takes the following form: 
\be
\la0|e^{i\a N(x)}|0\ra=\prod_{i=1}^{M}\int_{0}^{L}dx_i 
e^{i\a\sum_{i=1}^{M}\theta(x-x_i)}|\psi(x_1,x_2,...x_M)|^2, 
\label{exp}
\ee
where the ground state wave function is normalized to unity. 
The correlators for the inpenetrable Bose gas system and the 
free-fermion system are equal to each other. 
First consider the case $\a<\pi$. 
As in the case of the one-particle density matrix, we represent 
the correlator as the Toeplitz determinant: 
\be
\la0|e^{i\a N(x)}|0\ra=\det_{M}M_{ij}[f(y)], 
\label{det}
\ee
where the function 
$f(y)=e^{i2\pi b}\theta(2\pi x/L-y)+\theta(y-2\pi x/L)$, 
$b=\a/2\pi$, The representation of this function in the form 
(\ref{f}) is 
\[
f_{0}(y)=e^{ibx_r}=e^{i\a x/L}, ~~~x_1=0, ~~a_1=0, ~~b_1=-b, ~~
x_2=x_r, ~~a_2=0, ~~b_2=b, 
\]
where $x_r=2\pi x/L$. Substituting this function into the equations 
(\ref{DN}), (\ref{E}) we obtain the following result: 
\[
\la0|e^{i\a N(x)}|0\ra=e^{i\a Mx/L}\frac{1}{(2M\sin(\pi x/L))^{2b^2}} 
G(1+b)G(1-b). 
\]
Next, consider the case $\a=\pi$.  
As for the correlator $G_{\a}(x)$, at $\a=\pi$ ($b=1/2$) we have 
two different ways to represent the function $f(y)$ (\ref{det}) 
in the form (\ref{f}). Taking the sum of the two terms we obtain: 
\[
\la0|e^{i\pi N(x)}|0\ra=\frac{2\cos(\pi Mx/L)}{(2M\sin(\pi x/L))^{1/2}} 
G(3/2)G(1/2), 
\]
which is in fact a real function of $x$.

Finally, we consider the asymptotics of the density-density correlation 
function $\Pi(x)=\la0|\rho(x)\rho(0)|0\ra$, where $\rho(x)$ is the 
density operator. The expression of this correlator in terms of the 
ground state wave function reads: 
\be
\Pi(x)=M^2\prod_{i=3}^{M}\int_{0}^{L}dx_i |\psi(x,0,x_3,...x_M)|^2. 
\label{P}
\ee
Clearly this correlator is the same for the inpenetrable Bose gas system 
and the free fermion system. 
Performing the calculations we obtain the following expression of 
(\ref{P}) as a Toeplitz determinant: 
\[
\Pi(x)=\frac{4}{L^2}\sin^2(\pi x/L)\det_{M-2}M_{ij}[f(y)], 
\]
where the function $f(y)$ is represented in the canonical form (\ref{f}) 
and corresponds to the values $a_1=a_2=1$, $b_1=b_2=0$. 
Calculating the determinant we obtain 
\[
\det_{M-2}M_{ij}[f(y)]=\frac{(M-2)^2}{4\sin^2(\pi x/L)} 
\]
and the leading order term in the asymptotics of the density-density 
correlation function in the thermodynamic limit is 
 $\Pi(x)=\rho^2$, where the density $\rho=M/L$. 
Thus the generalized Fisher-Hartwig conjecture gives the correct 
result in this case too.

In conclusion, we have calculated both the exponential correlators 
and the usual ones in the one- dimensional Bose gas and the 
free-fermion systems. In the cases when the results can be obtained 
by the different methods, namely for the one-particle density matrices 
for the Bose and Fermi - systems, the predictions of the generalized 
Fisher-Hartwig conjecture \cite{BM} provide the additional tests 
of this conjecture. 
In the cases of various exponential correlators the predictions 
may be usefull in the context of the different approaches \cite{H}, 
\cite{C} to the calculation of the asymptotics of the correlation functions.

\end{document}